\documentclass[a4paper]{elsarticle}
\usepackage{lineno,hyperref}

\usepackage[utf8]{inputenc}
\usepackage{bm}
\usepackage{amsmath}
\usepackage{appendix}
\usepackage{xcolor}
\usepackage{amssymb}
\usepackage{soul}

\usepackage{nomencl}
\makenomenclature
\journal{Journal of \LaTeX\ Templates}

\begin{document}

\begin{frontmatter}


\title{Computational Fluid Dynamics and Machine Learning as tools for Optimization of Micromixers geometry}


\author[1,2]{Daniela de Oliveira Maionchi\corref{mycorrespondingauthor}}
\cortext[mycorrespondingauthor]{Corresponding author}
\ead{dmaionchi@fisica.ufmt.br}
\author[2]{Luca Ainstein}
\author[2]{Fabio Pereira dos Santos}
\author[2]{Maurício Bezerra de Souza Júnior}

\address[1]{Instituto de F\'isica , Universidade Federal de Mato Grosso - UFMT, 78060-900, Cuiab\'a-MT, Brazil}
\address[2]{Escola de Qu\'imica, Universidade Federal do Rio de Janeiro - UFRJ, 21941-909, Rio de Janeiro-RJ, Brazil}

\begin{abstract}
Microfluidic devices have become a new trend in different fields and have attracted attention due to their compact size and capability to deal with a small amount of fluid. Micromixing is an efficient way to mix a small amount of miscible fluids at this microfluidic level.
This work explores a new approach for optimization in the field of microfluidics, using the combination of CFD (Computational Fluid Dynamics), and Machine Learning techniques. 
The objective of this combination is to enable global optimization with lower computational cost. 
The initial geometry is inspired in a Y-type micromixer with cylindrical grooves on the surface of the main channel and obstructions inside it. 
Simulations for circular obstructions were carried out using the OpenFOAM software to observe the influences of obstacles.
The effects of obstruction diameter (OD), and offset (OF) in the range of $[20,140]$ mm and $[10,160]$ mm, respectively, on percentage of mixing ($\varphi$), pressure drop ($\Delta P$) and energy cost ($\Delta P/\varphi$) were investigated.
Numerical experiments were analyzed using machine learning. 
Firstly, a neural network was used to train the dataset composed by the inputs OD and OF and outputs $\varphi$ and $\Delta P$. 
The objective functions (ObF) chosen to numerically optimize the performance of micromixers with grooves and obstructions were $\varphi$, $\Delta P$, $\Delta P/\varphi$.
The genetic algorithm obtained the geometry that offers the maximum value of $\varphi$ and the minimum value of $\Delta P_s$. 
The results show that $\varphi$ increases monotonically with increasing OD at all values of OF. 
The inverse is observed with increasing offset. 
Furthermore, the results reveal that $\Delta P$ e $\Delta P/\varphi$ also increase with OD. 
On the other hand, the pressure drop and the cost of mixing energy present a maximum close to the lowest values of OF. 
Finally, the optimal value obtained for the diameter was OD$=131$ mm and for the offset OF$=10$ mm, which corresponds to obstruction of medium size close to the channel wall.
It is worth to mention that each simulation takes around $4h$, the total time to guarantee the global optimization would be about 330 days. With this methodology, the whole process of producing the dataset, training and optimization takes 40 days. This procedure is a tremendous advantage for microfluidic optimization.

\end{abstract}

\end{frontmatter}

\section{Introduction}

 Miniaturization of equipment has become necessary to develop engineering technologies to solve a series of problems \cite{capreto2011}.
With these technologies, besides enabling the integration of various functionalities on a single chip, called a Lab-on-chip (LOC), it is possible to reduce energy consumption, size of equipment/product factories, production capacity, and waste generation \cite{stank2000}.
The LOC has been used in several areas, such as nanoparticle crystallization, \cite{sta2001}, extraction, \cite{sprogies2008}, polymerization, \cite{nagaki2004}, organic synthesis, \cite{haswell2001}, enzyme assay , \cite{miller2008}, protein folding, \cite{bilsel2005}, bioprocess optimization, \cite{micheletti2006} and drug production studies, \cite{zafar2007}.

Computational Fluid Dynamics (CFD), an area that is concerned with combining physical knowledge about fluid mechanics with mathematical and computational resources and tools to predict, model and optimize flow parameters, plays a prominent role in several areas due to its versatility \cite{khan2018}.
Among these areas, we can mention its use in the study of atmospheric movements, \cite{mirzaeo2021}, biomedicine, \cite{wusten2021}), aerospace, \cite{wang2013}, oil and gas, \cite{martinez2020}), industrial equipment design (\cite{negi2021}, ocean engineering, \cite{forou2021} and microfluidics, \cite{chen2020, ortega2017}.

Understanding the relevant physical parameters in the analysis of micromixers occurs through the use of CFD, as it is not possible to treat the micromixer as an ordinary small-size mixer because this change in dimensions alters the physics involved once it creates a preferably laminar flow \cite{capreto2011}.
This paradigm shift makes the preferential turbulent mixing model present in regular mixers give way to a molecular diffusion model \cite{beebe2020}.
The challenge that arises with this physical model is that, to improve the mixing process, caps, \cite{wang2012} or obstructions, \cite{afroz2014} are added in T, \cite{nimafar2012,zhendong2012}, H, \cite{nimafar2012}, O, \cite{nimafar2012}, Y, \cite{wang2012} geometries or fractals, \cite{chen2020}, end up increasing the pressure drop, which directly influences the mixing energy cost, \cite{rahma2019}.

In this sense, equipment optimization involves the determination of geometry, flow rates and pressure drop analysis \cite{wang2012}.
The main issue encountered in the optimization process is the generation of data. If an experimental procedure is used, it is necessary to build several geometries and test in different configurations, which generates a high cost of experimentation.
Computational calculation approach, using CFD, requires time to generate a relevant number of simulations, in addition to having a computational cost.
This makes the optimization work, both experimental and simulation, to have a limited number of test cases \cite{wang2012,nimafar2012,rahma2019}.
To avoid this difficulty, Machine Learning appears as an intelligent alternative, as it is a type of technology that learns from the generated data and allows predictions to be made in a much smaller time window than that required by the techniques of existing CFDs.

%
%

Although several works combine CFD with machine learning, the literature review showed that few articles use machine learning in microfluidics. %
Of these few works, it is worth mentioning the article of, \cite{hadik2019,QUEIROZ2021100002}, where a large number of droplet images are recorded and used to train deep neural networks (DNN) to predict flow or concentration.
It is shown that this method can be used to quantify the concentrations of each component with an accuracy of $0.5\%$ and the flow rate with a resolution of $0.05$ ml/h.
\cite{arjun2020} performed a detection and classification of coalesced binary drops within microchannels based on the degree of mixing using a deep neural network.
Hence, the use of machine
learning to reduce order in optimization problems employing CFD in micromixers has no precedent in the literature. 

In section \ref{model}, the micromixer modelling as developed by \cite{wang2012, rahma2019} is presented, together with the results obtained in these articles.
Results and discussions are shown in section \ref{results}, including model validation, neural network training and test, as well as optimization and verification of optimal values.
Finally, the conclusion summarizes the goals achieved and future proposals.

\section{Methodology}\label{model}

This work is based on the articles by \cite{rahma2019, wang2012} and aims to optimize one of the geometries proposed in \cite{rahma2019} using Machine Learning, more specifically dense feedforward neural networks for deriving a reduced-order model and genetic algorithm to obtain a global minimum.

As specific objectives, it is intended to obtain the dimensions of circular obstacles (diameter and offset) included along the micromixer channel that maximize the mixing of two liquids with different concentrations and minimize the head loss in the channel.
The tests performed correspond to the construction of the system's Pareto curve.
The solutions obtained by this method for different points on the curve, are applied in simulations to verify the values found.

\begin{figure}[h!]
\centering
\includegraphics[width=0.9\textwidth]{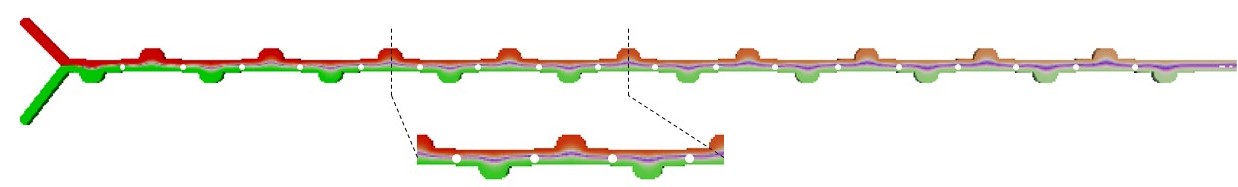}
\caption{Schematic diagram of the micromixer.}
\label{Fig2_modelo}
\end{figure}

Mixing is fundamentally important in microfluidic systems.
For this work, a Y-geometry micromixer was selected (Figure \ref{Fig2_modelo}), with the dimensions described in Table \ref{Tabela1}, \cite{wang2012, rahma2019}.

\begin{figure}[h!]
\centering
\includegraphics[width =0.8\textwidth]{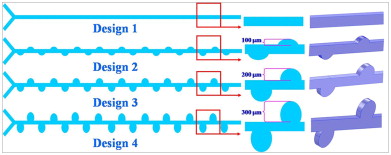}
\caption{Two-dimensional (top view) and three-dimensional (right column) scheme of the four designs studied in \cite{wang2012}. Design 1 is the straight channel and designs 2, 3, and 4 are the channels with CGs $1/4$, $1/2$ and $3/4$, respectively.}
\label{figtri}
\end{figure}

Three simple but innovative passive mixers with cylindrical grooves (CG) adjacent to the main channel were designed and manufactured in \cite{wang2012}, as shown in Figure~\ref{figtri}. 
A series of simulations and experiments were carried out for different groove depths, ranging from $0$ to $3/4$ cylindrical groove.
All results confirmed the improvement of mixing in these mixers over a short distance ($< 20 mm$).

\begin{table}[t]
\centering
\caption{Dimensions of micromixer.}
\begin{tabular}{l l}
\hline
Channel lenght (L) & $2cm$\\
Channel width (W) & $200\mu m$\\
Channel depth (d) & $20 \mu m$\\
Distance between twoo CGs & $1000 \mu m$\\
Diameter of CGs & $200 \mu m$\\
Distance between two obstructions & $1000 \mu m$\\
\hline
\end{tabular}
\label{Tabela1}
\end{table}

Computational fluid dynamics (CFD) and response surface methodology (RSM) were used in \cite{rahma2019} to optimize slotted micromixers with obstructions.
The initial geometry was inspired by \cite{wang2012}, with obstructions in the form of a circle.

The effects of occlusion dimension (OD) and displacement (OF) were investigated on mixing percentage, $\varphi$, pressure drop, $\Delta P$, and mixing energy cost, \emph{mec}.
%
%
The results of \cite{rahma2019} showed that $\varphi$ increases monotonically with the
increase in OD at all OF values for all forms of obstruction (circle, square and diamond).
The inverse trend was also observed with increased displacement.
Furthermore, the results revealed that $\Delta P$ and \emph{mec} increase with increasing OD at almost all OF values for all forms of obstructions.
On the other hand, the pressure drop and the cost of mixing energy reduce as OF increases.
$\varphi$, $\Delta P$ and \emph{mec} were considered as the objective functions to numerically optimize the performance of slotted grooved micromixers with obstructions.
%
%


In the present work, the optimization of the micromixer geometry was performed using Machine Learning, more specifically neural networks and genetic algorithm.
All simulations were performed using OpenFOAM for stationary and incompressible flow using the finite element method.
The obstruction of channels with cylindrical grooves CG = $1/2$ was considered to be in a circle shape, in a range of OD obstruction dimensions from $20$ to $140 \mu m$ and OF obstruction displacement from $10$ to $160 \mu m $ (the offset in this case refers to the distance from the bottom point of the circle to the bottom wall of the channel for all possible OD values).
%

The continuity, Navier-Stokes and convection-diffusion-species equations (Eq. \ref{eq1_model}) were solved for a uniform laminar flow of a Newtonian fluid with constant properties.
The flow has the same velocity at both inputs, but different mass concentrations of solute $C$, being
$C = 1 \quad mol/ m^3$ in the lower entry and $C = 0 \quad mol/ m^3$ in the upper entry, ie, free of solute.

\begin{eqnarray}
     \overrightarrow{\nabla}.\overrightarrow{u} &=&0, \\
     \rho \frac{D\overrightarrow{u}}{Dt}&=&- \overrightarrow{\nabla}p+\mu \nabla^{2} \overrightarrow{u}, \\
     \frac{DC}{Dt}&=&\gamma \nabla^{2} C,
\end{eqnarray}\label{eq1_model}
where $\overrightarrow {u}$, $p$ and $C$ are, respectively, the velocity, pressure and concentration, and $\rho$, $\mu$ and $\gamma$ are, respectively, the density, viscosity and fluid diffusion coefficient.

The boundary conditions used in the inputs were normal uniform velocity, based on the Reynolds number, no slip on the walls and gauge pressure ($p = 0 Pa$) at the output.
The density, dynamic viscosity and diffusion coefficient of the fluid are $\rho = 998kg/m^3$,$\mu =8.9 \cdot 10^{-4} Pa \cdot s$ and $\gamma = 10^ {-9} m^2/s$, respectively.
The Reynolds number is defined as $Re = \rho U_{in} W/l$, where $U_{in}$ is the flow velocity at the inputs
and $W$ is the channel width.
The Reynolds number was considered $Re \approx 1$ in the validation case and in the
other simulations.

The distribution of concentration levels across the width of the main channel can be used to assess the level of fluid mixing in micromixers.
The blend percentage ($\varphi$) is determined by the following equation, \cite{lin2007},

\begin{equation}
    \varphi = \left(1-\frac{\sigma}{\sigma_{max}} \right )\cdot 100\%, \label{phi}
\end{equation}
where $\sigma$ is the standard deviation, and the subscript \emph{max} denotes the initial unmixed state in the micromixer ($0.5$ in this case).
The standard deviation can be calculated by the concentration distribution as

\begin{equation}
     \sigma = \sqrt{\frac{1}{N-1}\sum^{N}_{i=1}(C_{i}-\overline{C_{i}})^2}, \quad \overline{C_{i}} = \frac{\sum_{i=1}^N C_i}{N},
\end{equation}
where $N$ is the number of sampling across the channel width, $C_i$ is the concentration of sampling i, and $\overline{C_i}$ is the mean value of the concentration.
In addition, the mixing energy cost (\emph{mec}), \cite{casanova2017}, is also used to estimate the efficiency of the micromixers, as it measures the pumping power needed to obtain one percent of the mixture.
So it can be defined as

\begin{equation}
     mec = \frac{Q\Delta P}{\varphi},
     \label{eq2_model}
\end{equation}
where $Q$ is the flow rate through the mixing channel and $\Delta P$ is the pressure difference between the output and inputs of the channel.
As $Q$ is constant, in this work, we will consider only the value of the ratio $\Delta P/\varphi$.

The governing equations were solved until the residuals reached below $10^{-9}$, which means that all flow properties remain constant throughout the iterations.
The methodology of this study is validated by reproducing the results of \cite{wang2012, rahma2019}.
The "design 2" presented in Figure \ref{figtri} is considered with the same parameters as in \cite{wang2012}.

Figure \ref{Fig2_modelo} represents the geometry of the CG $1/2$ used for geometries with obstructions present.
However, to compare the simulation with the experimental data, a geometry without obstructions and with CG $1/4$ was adopted.
Similar to what was done in \cite{rahma2019}, the validation of the model used in this work was done by comparing the results obtained for $\varphi$ experimentally and with OpenFOAM and are presented in the next section.

In order to apply neural networks to the micromixer with circular obstructions, several simulations with different values of OD and OF, and consequently $\varphi$ and $\Delta P$, must be carried out, in order to obtain enough data to be used for training and test.
Again, it is worth emphasising the care so that there is no overfitting, as this can affect the validation of the obtained network.

Then, in possession of the model that best describes the fluid dynamics of the micromixers, the genetic algorithm method were used to optimize this model. 
This step basically consists of finding the maximum or minimum point of a function, which will be done in the model obtained for $\varphi$, $\Delta P$ and $\Delta P/\varphi$.

\section{Results}\label{results}

We performed $265$ simulations, changing the $OD$ and $OF$ values, in an HPC Cluster with processor Intel
Xeon\textsuperscript{\textregistered} E5-2640 v4 2.4GHz, where the training
calculation was performed mainly in a Tesla P100 GPU with 16GB VRAM.
In this case, each simulation of our dataset, took around $4$ hours.

Figure \ref{Fig2_modelo} illustrates one of the cases with simulated obstacles in OpenFOAM.
The simulation results were used to train a dense neural network using the TensorFlow library. %
This neural network can predict new configurations (changing OD and OF) that can be used to carry out the global optimization process.
Therefore, the neural network was trained considering the values of $OD$ and $OF$ as input variables and the values of $\Delta P/\varphi$ and $\varphi$ as output.
With an adequate neural network model, the optimization was performed in order to maximize $\varphi$ and minimize $\Delta P/\varphi$.
The GeneticAlgorithm and Platypus Python libraries were used for the cases of single objective and multiobjective functions, respectively.

\subsection{Validation}

\begin{figure}[h!]
\begin{center}
\resizebox*{\textwidth}{!}{
\includegraphics[width=\textwidth]{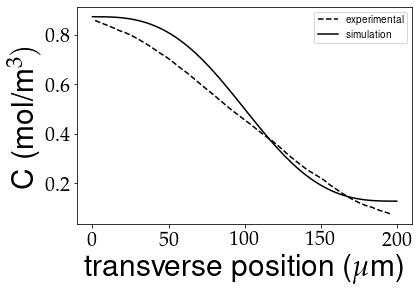}
\includegraphics[width=\textwidth]{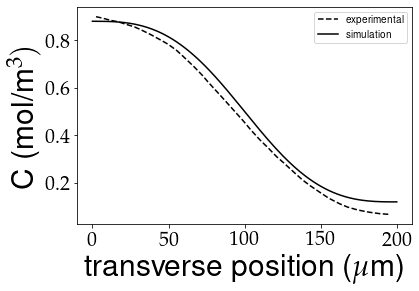}
}
(a)\hspace{5.5cm} (b)
\caption{Comparison between experimental and numerical (present study)
concentration distribution at the distance of 16 mm for CG (a) $1/4$ and (b) $1/2$, with $Re \approx 1$.}
\label{comparison}
\end{center}
\end{figure}

The concentration distribution at the position $x = 16mm$ was numerically evaluated in this work and compared with the experimental results presented in \cite{wang2012, rahma2019}, as shown in Figure \ref{comparison}.
The concentration distribution is related to the percentage of the mixture ($\varphi$).
The results obtained show that the simulation results are consistent with the experimental results.

\begin{figure}[h]
\begin{center}
\resizebox*{\textwidth}{!}{
\includegraphics[width=\textwidth]{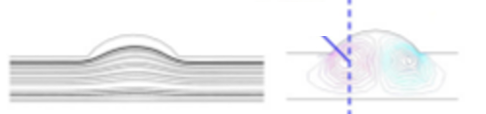}
\includegraphics[width=\textwidth]{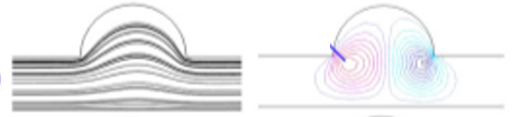}}
(a)\hspace{5.5cm} (b)
\caption{Simulated results obtained by \cite{wang2012} of streamlines and transverse velocity contour of designs CG (a) $1/4$ and (b) $1/2$, with $Re = 1$.}
\label{concentration}
\end{center}
\end{figure}

\begin{figure}[ht!]
\begin{center}
\resizebox*{\textwidth}{!}{
\includegraphics[width=\textwidth]{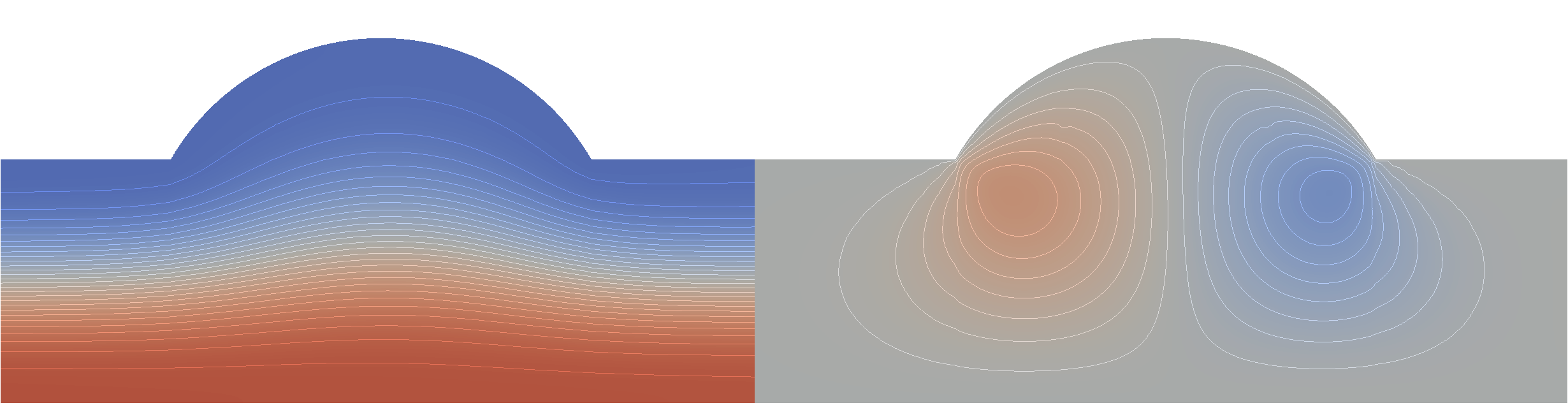}}
\resizebox*{\textwidth}{!}{
\includegraphics[width=\textwidth]{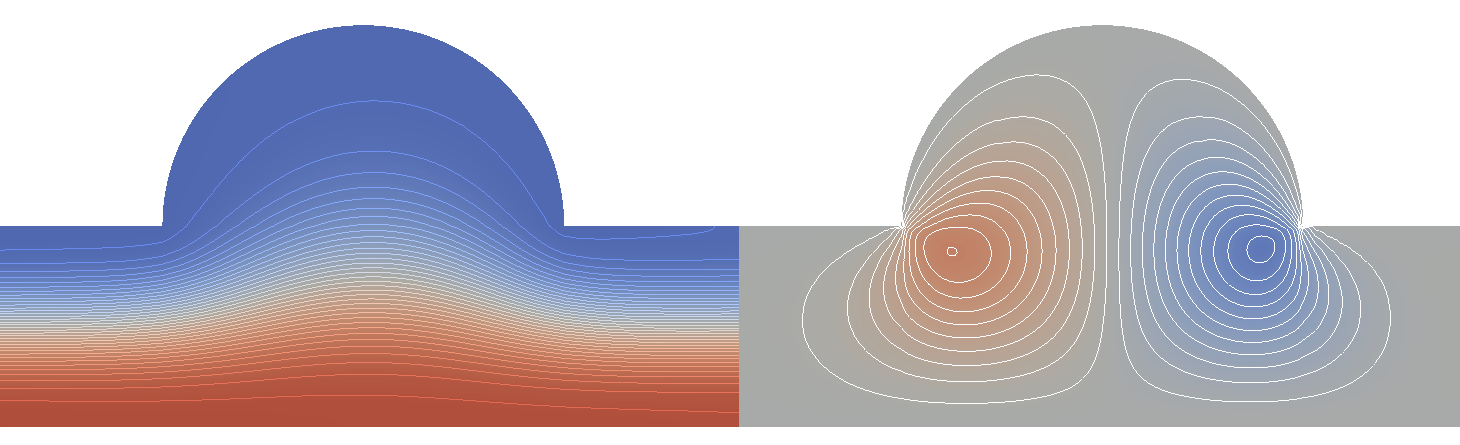}}
\caption{Simulated results obtained with OpenFOAM of streamlines and transverse velocity contour of designs CG $1/4$ (top) and $1/2$ (bottom), with $Re \approx 1$.}\label{fig01}
\label{streamlines}
\end{center}
\end{figure}

In addition, the comparison between the flow lines and the transverse velocity in the cap region was also performed.
As can be seen in Figures \ref{concentration} and \ref{streamlines}, the results obtained computationally showed again a lot of precision in relation to the experimental data.

The data in Figure~\ref{comparison} can be used to evaluate the predicted mixture percentage experimentally (\cite{wang2012}) and numerically (this study).
Using Eq. \ref{phi}, $\varphi$ is calculated for CG $1/4$ and $1/2$, respectively, as $45.5\%$ and $39.1\%$ by the experimental data from \cite{wang2012}, and $42.4\%$ and $41.4\%$ for the numerical data for this study.
Considering the methodological differences between experiment and simulation, the agreement between the results is considered good enough.

\begin{figure}[hb!]
\begin{center}
\resizebox*{\textwidth}{!}{
\includegraphics[width=0.48\textwidth]{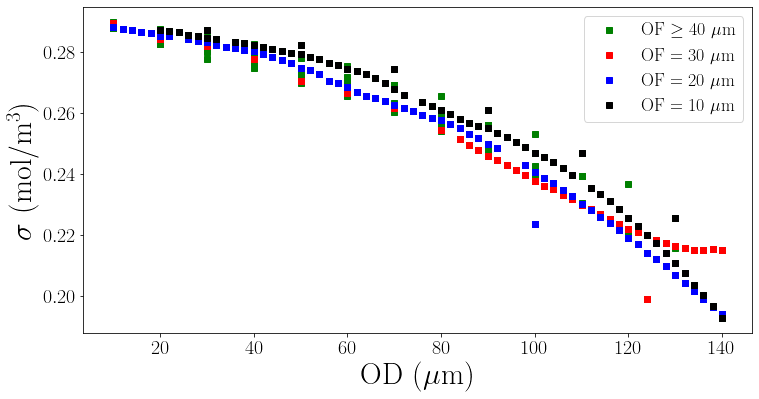}
\includegraphics[width=0.48\textwidth]{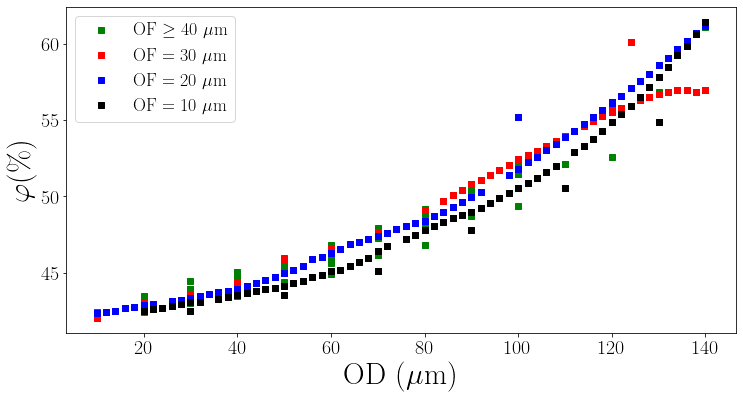}
}
\resizebox*{\textwidth}{!}{
\includegraphics[width=0.48\textwidth]{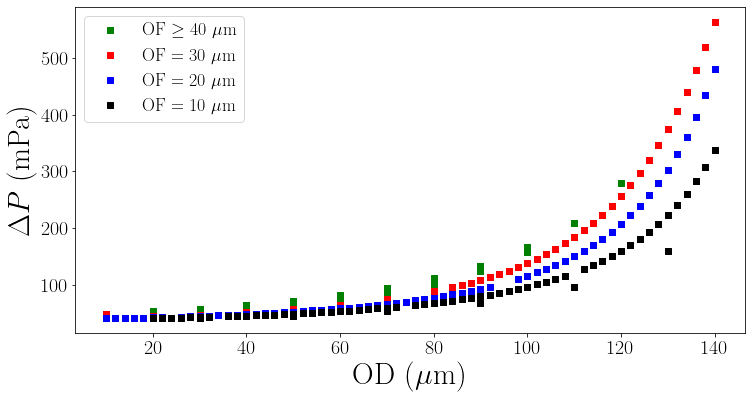}
\includegraphics[width=0.48\textwidth]{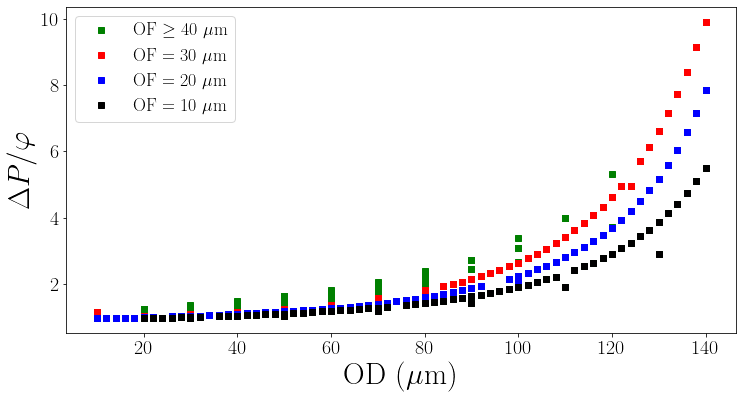}
}
\caption{Data obtained from simulations: $\sigma$, $\varphi$, $\Delta P$ e $\Delta P/\varphi$ as function of  OD.}\label{fig6_resultado}
\label{var_OD}
\end{center}
\end{figure} 

\begin{figure}[ht!]
\begin{center}
\resizebox*{\textwidth}{!}{
\includegraphics[width=0.48\textwidth]{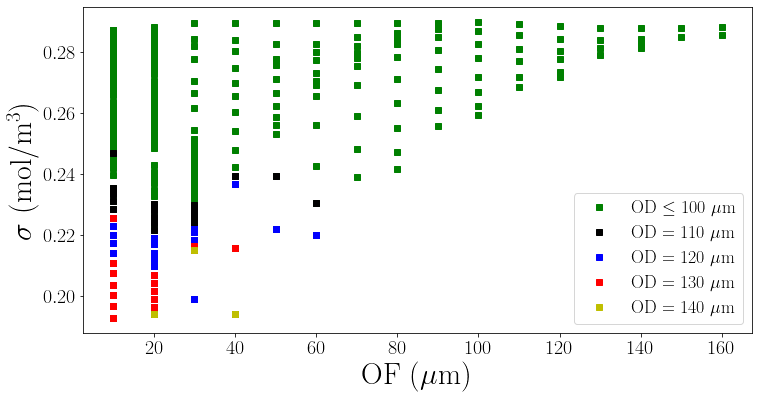}
\includegraphics[width=0.48\textwidth]{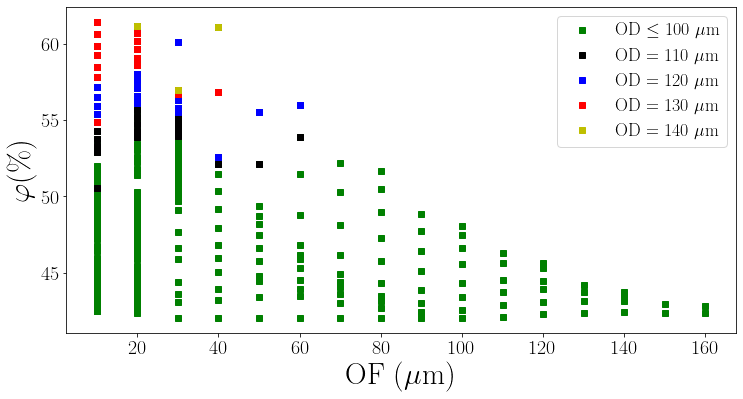}
}
\resizebox*{\textwidth}{!}{
\includegraphics[width=0.48\textwidth]{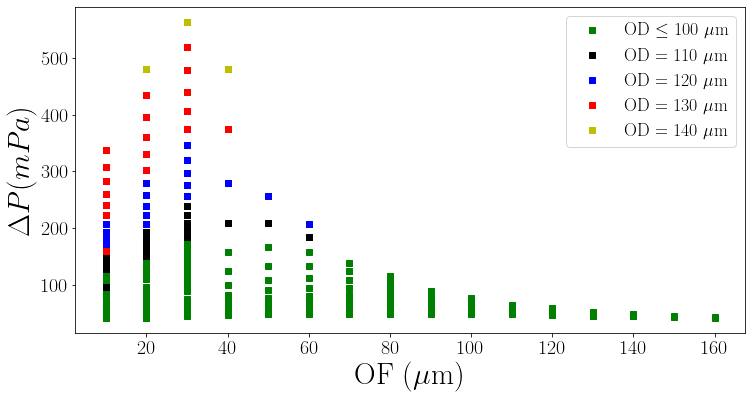}
\includegraphics[width=0.48\textwidth]{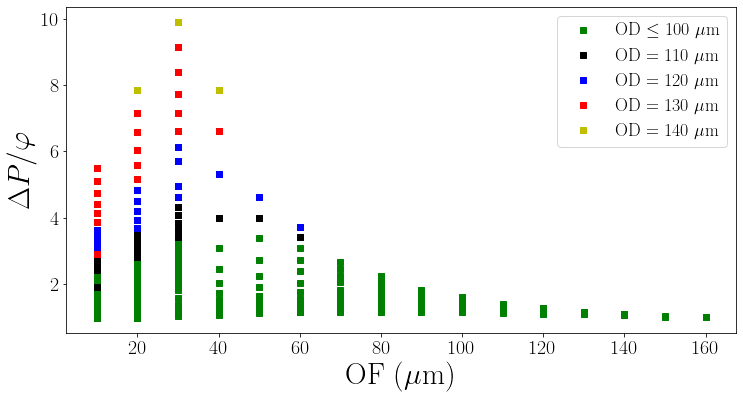}
}
\caption{Data obtained from simulations: $\sigma$, $\varphi$, $\Delta P$ e $\Delta P/\varphi$ as function of  OF.}
\label{var_OF}
\end{center}
\end{figure} 

\begin{figure}[hb!]
\begin{center}
\resizebox*{\textwidth}{!}{
\includegraphics[width=0.98\textwidth]{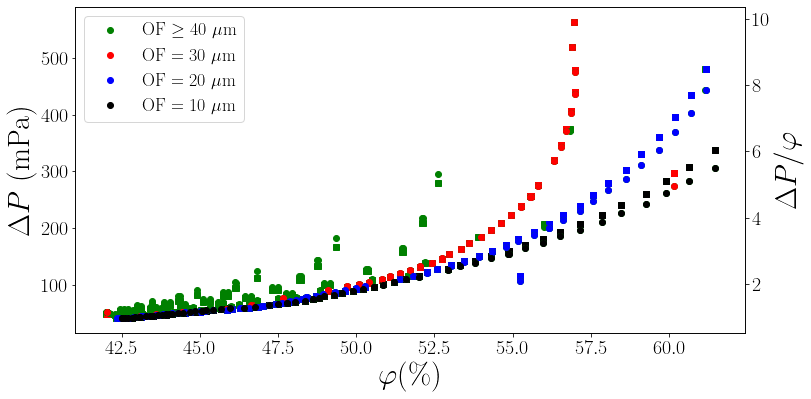}
}
\caption{Pressure drop $\Delta P$ and $\Delta P/\varphi$ as functions of the mixture $\varphi$.}
\label{Pversusphi}
\end{center}
\end{figure} 

From the results obtained from the simulations in the distribution through the cross section of the micromixer in the position $16 mm$, it was possible to define $\sigma$, $\varphi$ and $\Delta P/\varphi$, shown in Figs. \ref{var_OD} and \ref{var_OF}.
The values obtained in the simulation performed were compared with those described in the literature, being consistent with what was expected.

Interestingly, as OD increases, both $\varphi$, $\Delta P$ and $\Delta P/\varphi$ also increase. As OF increases, both $\varphi$, $\Delta P$ and $\Delta P/\varphi$ tend to decrease.
This shows how difficult it is to find OD and OF values that simultaneously satisfy the maximization of $\varphi$ and the minimization of $\Delta P$ and $\Delta P/\varphi$.
Figure~\ref{Pversusphi} shows the relationship between $\Delta P$ and $\varphi$ that must be optimized.

\subsection{Neural network training and test}

Part of the results shown in Figures~\ref{var_OD} and \ref{var_OF} were used to train a dense neural network, which generated a predictive model that allows, from a given geometry, to predict the values of the pressure drop $ \Delta P$, the mixing percentage $\varphi$ and $\Delta P/\varphi$.

The data obtained in the simulations were organized containing the variables to be used in the neural network, that is, OD and OF for input, $\varphi$ and $\Delta P$ for output.
These data were separated into training data and test data with the ratio value  $0.2$, which means that $20\%$ of the data was be used for test and $80\%$ for training.
Herein, we mapped values in the range from $0$ to $1$, which ensures that the weight of all variables, both input and output, are the same.

The neural network requires the definition of the model, we defined our dense neural network.
It is worth emphasizing here the importance of observing through the graph whether overfitting occurs, which is identified by the training data presenting a value higher than the curve of the simulation data. 
Finally, test data can be applied to the found network for its validation.

\begin{figure}[h]
\begin{center}
\resizebox*{\textwidth}{!}{
\includegraphics[width=\textwidth]{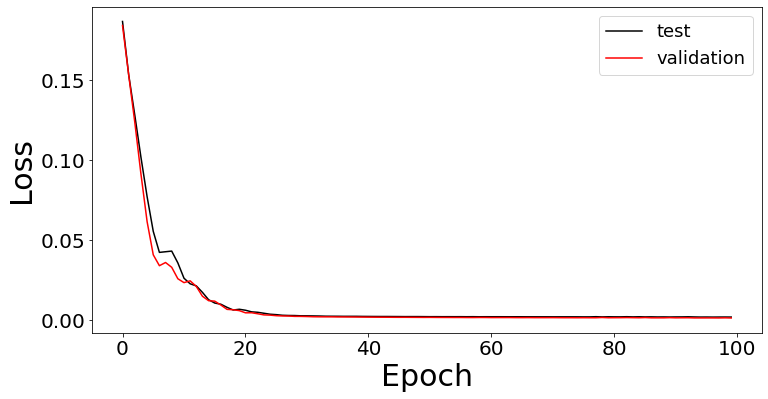}
\includegraphics[width=\textwidth]{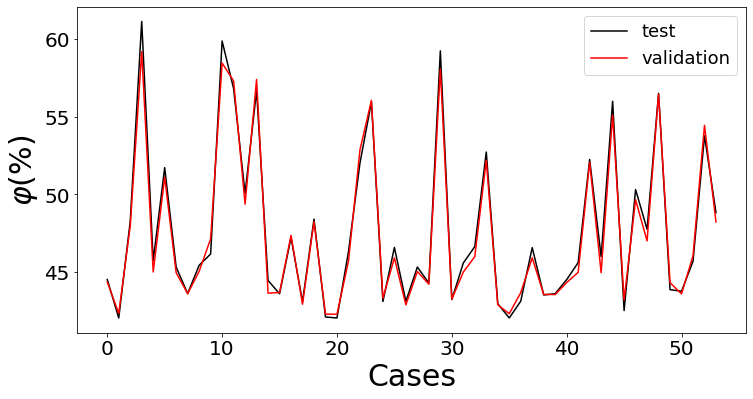}
}
(a) \hspace{5.5cm} (b)
\caption{(a) Loss function for the training data and (b) comparison of the value $\varphi$ between the test and validation data. The average error calculated was $0.979\%$.}
\label{rede1}
\end{center}
\end{figure}

\begin{figure}[h]
\begin{center}
\resizebox*{\textwidth}{!}{
\includegraphics[width=\textwidth]{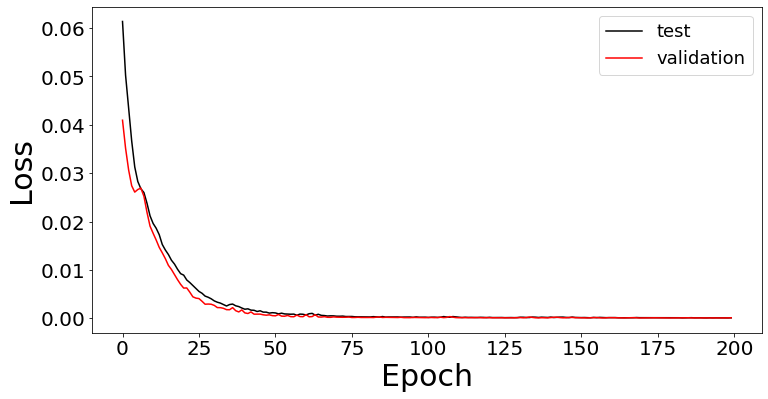}
\includegraphics[width=\textwidth]{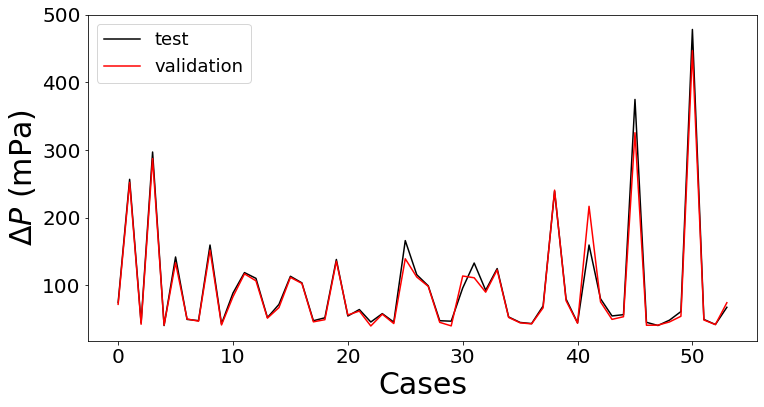}
}
(a) \hspace{5.5cm} (b)
\caption{(a) Loss function for the training data and (b) comparison of the value $\Delta P$ between the test and validation data. The average error calculated was $2.604\%$.}
\label{rede2}
\end{center}
\end{figure}

\begin{figure}[h]
\begin{center}
\resizebox*{\textwidth}{!}{
\includegraphics[width=0.5\textwidth]{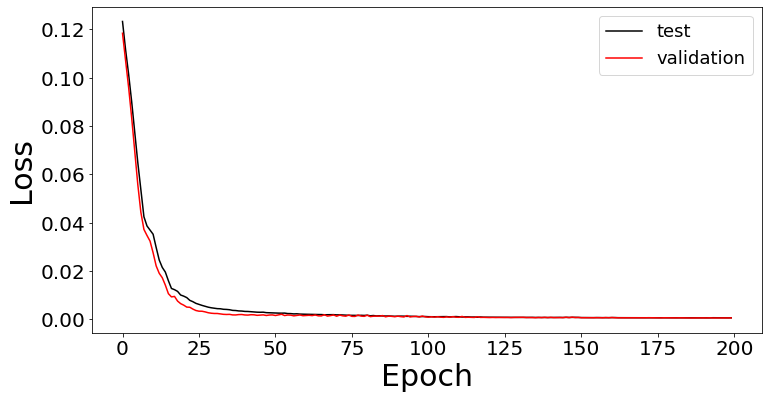}
\includegraphics[width=0.5\textwidth]{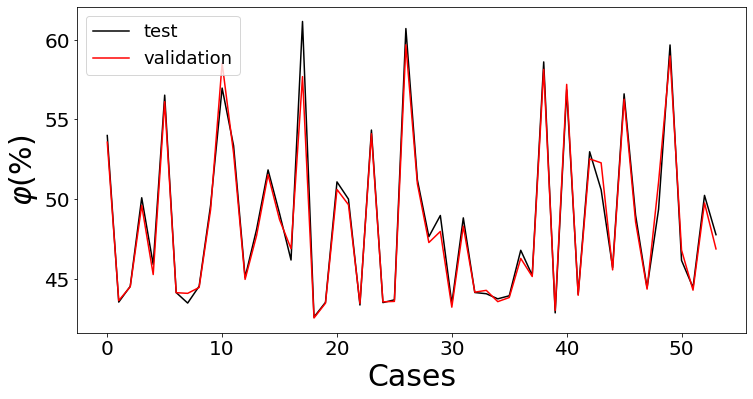}
}
(a) \hspace{5.5cm} (b)
\resizebox*{\textwidth}{!}{
\includegraphics[width=0.5\textwidth]{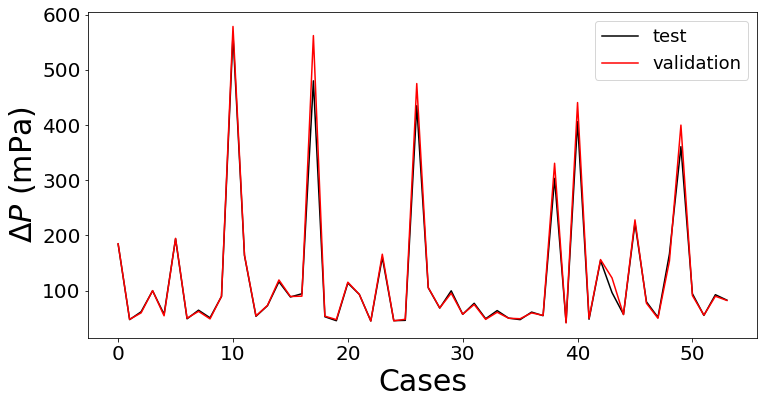}
\includegraphics[width=0.5\textwidth]{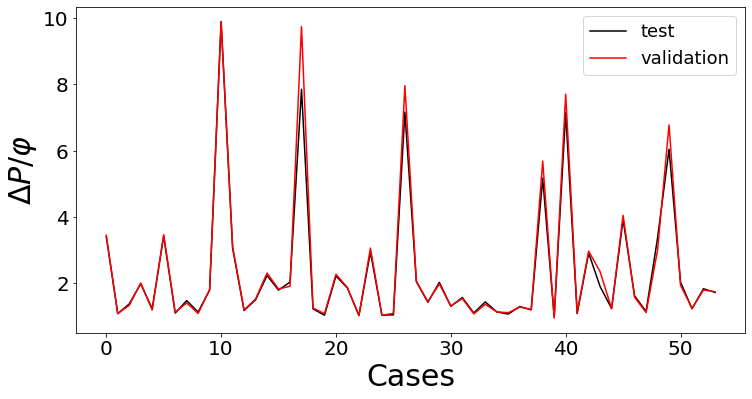}
}
(c) \hspace{5.5cm} (d)
\caption{(a) Loss function for the training data, comparison of (b) $\varphi$, (c) $\Delta P$ and (d) $\Delta P/\varphi$ between the test and validation data. The average error calculated was $0.972\%$ and $1.820\%$, respectively.}
\label{rede3}
\end{center}
\end{figure}

The mean error for pressure drop was less than $6 \%$, while the mean error for percentage $\varphi$ was less than $1.2\%$.
The comparison between the values obtained in the simulations using OpenFOAM (labeled as ''experimental") and those predicted by the neural network (labeled as "validation"), from the OD and OF values, are shown in Figures~\ref{rede1} - \ref{rede3}.


\subsection{Optimization and verification}

The optimization was first performed using Python's Geneticalgorithm library, which allows the definition of only one objective function.
Three different optimizations involving the maximization and minimization of (i) $\varphi$, (ii) $\Delta P$ and (iii) $\Delta P /\varphi$ were considered.

As an example, $\Delta P/\varphi$  was chosen as the objective function
with the developed network model, the input data is OD and OF and the prediction of these values are used as output of the objective function.
Furthermore, the function is subject to a restriction due to the width of the channel, where the diameter of the obstacle must be smaller than this value, that is, $OD + OF >200 \mu m$. If this relationship is satisfied, it must assign a penalty value that will be added to the output of the objective function.
The penalty should be a value much higher than those usually obtained at the exit to ensure that this situation is far from the minimum of the function.

If, instead of the global (or local) minimum, the interest is in finding the maximum of the objective function, the only change to be made is in the output signal turning it to minus.

Once the appropriate objective function is defined, the execution of the genetic algorithm is performed.
The parameters of this algorithm must be defined, such as population size, mutation and crossover rate  and maximum number of iterations.
In this problem we chose to exclude the cases with very low $\Delta P$, as they also correspond to low $\varphi$ ($< 50\%$), which is obtained when considering OD ranging only from $10$ to $150$ and OF from $10$ to $160$.

The values obtained for the objective functions, as well as the respective OD and OF are presented in Tables~\ref{Tabela2} and \ref{Tabela3}.

\begin{table}[h!]
\centering
\caption{Maximum values obtained with the optimization.}
\begin{tabular}{c c c c c c}
\hline
ObF & Maximum & $\varphi$ & $\Delta P$ & OD ($\mu$m) & OF ($\mu$m)\\
\hline
$\varphi$ ($\%$) & $59.45$ & $59.45$ & - & $139$ & $41$\\
$\Delta P$ (mPa) & $582.99$ & - & $582.99$& $144$ & $51$\\
$\Delta P/\varphi$ & $9.36$ & $59.59$ & $557.66$ & $142$ & $34$\\
$\Delta P_s+1/\varphi_s$ & $111.39$ & $42.69$ & $51.34$ & $102$ & $99$\\
\hline
\end{tabular}
\label{Tabela2}
\end{table}

\begin{table}[h!]
\centering
\caption{Minimum values obtained with the optimization.}
\begin{tabular}{c c c c c c}
\hline
ObF & Minimum & $\varphi$ & $\Delta P$ & OD ($\mu$m) & OF ($\mu$m)\\
\hline
$\varphi$ ($\%$)& $42.28$ & $42.28$ & - & $10$ & $76$ \\
$\Delta P$ (mPa) & $40.79$ & - & $40.79$ & $14$ & $16$\\
$\Delta P/\varphi$ & $0.96$ & $42.71$ & $41.05$ & $11$ & $11$\\
$\Delta P_s+1/\varphi_s$ & $1.65$ & $56.91$ & $224.12$ & $127$ & $12$\\
\hline
\end{tabular}
\label{Tabela3}
\end{table}

The maximum values found for OD are very similar for $\varphi$, $\Delta P$ and $\Delta P/\varphi$, that is, around $140 \mu m$ , while the values of OF are small, not exceeding $34 \mu m$.
That is, they represent a geometry with large obstacles and close to the channel wall.
For $\Delta P_s+1/\varphi_s$, the found OD value is smaller, $102 \mu m$, with OF $99 \mu m$.
This corresponds to geometries with diameters around half the width of the channel, which are also located close to the channel wall.
Note that in the latter case, the values for $\varphi$ and $\Delta P$ are much smaller than for $\Delta P/\varphi$ which results in $1.20$.

For the minimal cases, the OD values for $\varphi$, $\Delta P$ and $\Delta P/\varphi$ are also small, between $10$ and $14 \mu m$, while OF varies between $11$ and $76 \mu m$.
In other words, small obstacles close or not to the wall generate flow with very low values for both variables, what should be expected.
On the other hand, $\Delta P_s +1/\varphi_s$ presents a higher OD value, $127 \mu m$ for a small OF value, $12 \mu m$. Note that $\Delta P_s$ and $\varphi_s$ are normalized to stay between 0 and 1.
While the minimum value of $\Delta P/\varphi$ has low values for both $\Delta P$ and $\varphi$, the same does not occur for its maximum value, which indicates that this objective function can be used to obtain a condition of $\varphi$ high and $\Delta P$ as low as possible.

\begin{figure}[ht!]
\begin{center}
\resizebox*{\textwidth}{!}{
\includegraphics[width=0.98\textwidth]{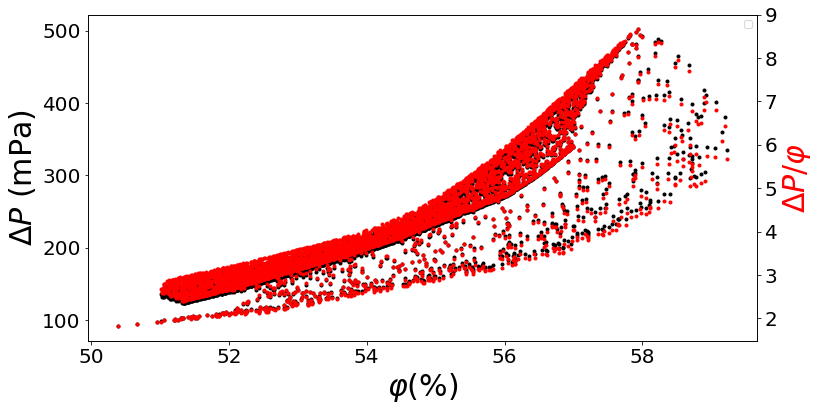}
}
\resizebox*{\textwidth}{!}{
\includegraphics[width=0.98\textwidth]{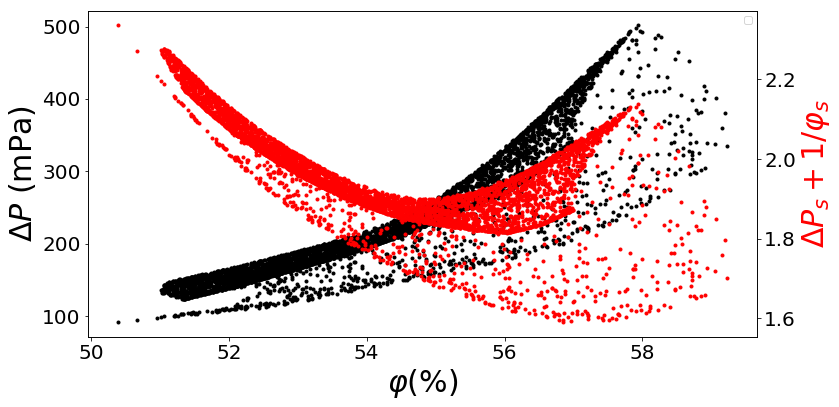}
}
\caption{Pareto curves that maximize $\varphi$ and minimize $\Delta P$ and $\Delta P/\varphi$ or $(\Delta P_s+1/\varphi_s)$.}
\label{pareto}
\end{center}
\end{figure}

These optimization results correspond to the adoption of a single objective function, $\varphi$, $\Delta P$ or $\Delta P/\varphi$.
Note that, according to Figures~\ref{var_OD} and \ref{var_OF}, both $\varphi$ and $\Delta P$ increase with the value of OD and decrease with the value of OF.
Thus, the objective of finding OD and OF values that simultaneously satisfy the conditions of $\varphi$ maximum, $\Delta P$ minimum and $\Delta P/\varphi$ minimum cannot be achieved as the method shown above.
It becomes necessary to use a multiobjective genetic algorithm, which was done using Python's Platypus library.
The three objective functions, in addition to these functions, must return the same restriction condition used previously.
One must inform the number of input variables, objective functions and constraints to the problem.
Furthermore, the restriction contained in the objective function output must be complemented so that the range of possible values for the input variables is well defined.
The genetic algorithm method used was NSGAII \cite{996017}.
At the end, the feasible solutions were filtered out of all the possible solutions found.
Pareto curves, as well as simulation data, are shown in Figure~\ref{pareto} for easy comparison.

The first curves shown in Figure~\ref{pareto} correspond to the best possible values of the objective functions, ranging from the prioritization of minimizing $\Delta P$ and $\Delta P/\varphi$ to maximizing $\varphi$.
The value found for $\varphi_{max}$ also corresponds to the highest value of $\Delta P$.
Likewise, the value found for $\Delta P_{min}$ also corresponds to the smallest value of $\varphi$.
In this work, we will consider, in addition to the optimal point obtained from $(\Delta P_s+1/\varphi_s)_{min}$, the points with $\Delta P/\varphi$ minimum and $\varphi$ maximum.
In Table~\ref{Tabela4}, the chosen case is summarized, as well as the results obtained in the verification simulation.

\begin{table}[h]
\centering
\caption{Comparison between the values obtained with optimization and simulation.}
\begin{tabular}{c c c c c c c}
\hline
ObF & OD ($\mu$m) & OF ($\mu$m) & $\varphi_{opt}$ & $\Delta P_{opt}$ (mPa) & $\varphi_{sim}$ & $\Delta P_{sim}$ (mPa)\\
\hline
$(\Delta P_s+1/\varphi_s)_{min}$ & $131$ & $10$ & $57.47$ & $235.87$ & $57.98$ &  $227.67$\\
\hline
\end{tabular}
\label{Tabela4}
\end{table}

From the OD and OF values obtained for $(\Delta P_s+1/\varphi_s)_{min}$ and observing Figures~\ref{var_OD} and \ref{var_OF}, it can be noted that the case Obtained refers to geometry close to the channel wall, with a medium-sized obstacle.
In this region, $\Delta P$ tends to decrease and $\phi$ remains large, as observed in \cite{rahma2019}.
Thus, the optimized geometry is compatible with what is expected, that is, obstacles that are not too big or too small and close to the channel wall should be prioritized.
 
\begin{figure}[h]
\begin{center}
\resizebox*{\textwidth}{!}{
\includegraphics[width=\textwidth]{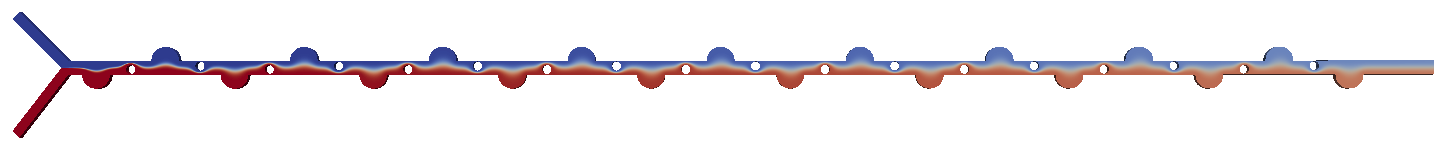}
}
\resizebox*{\textwidth}{!}{
\includegraphics[width=\textwidth]{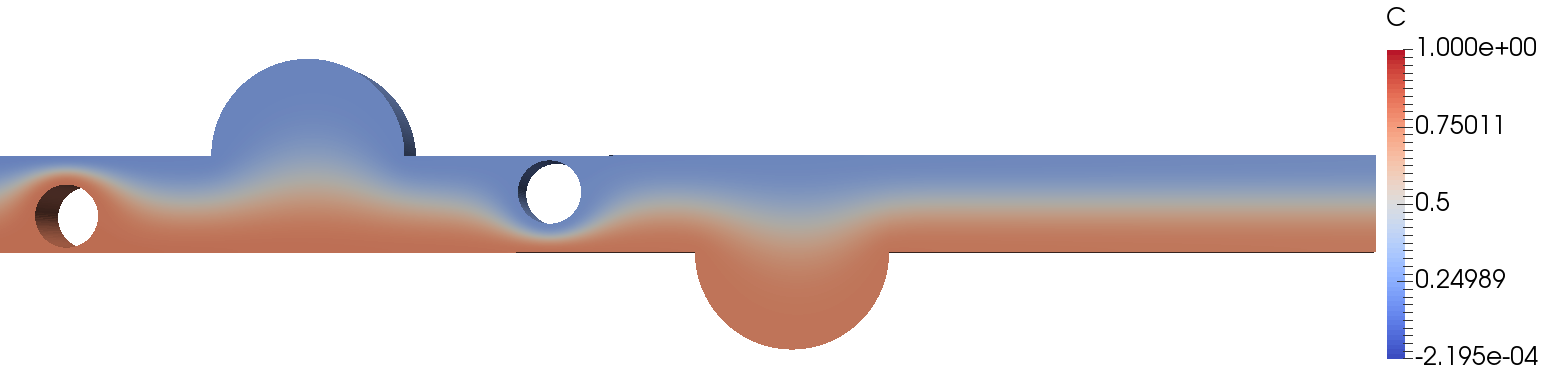}
}
\resizebox*{\textwidth}{!}{
\includegraphics[width=\textwidth]{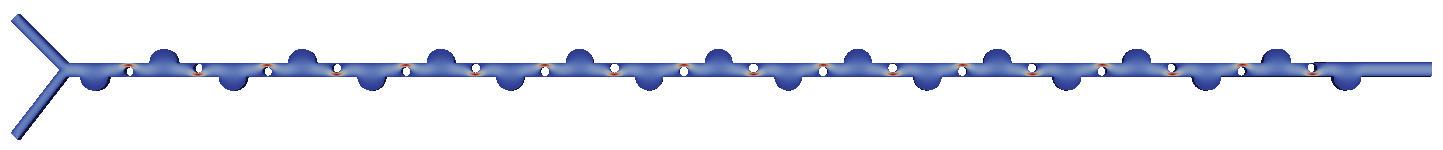}
}
\resizebox*{\textwidth}{!}{
\includegraphics[width=\textwidth]{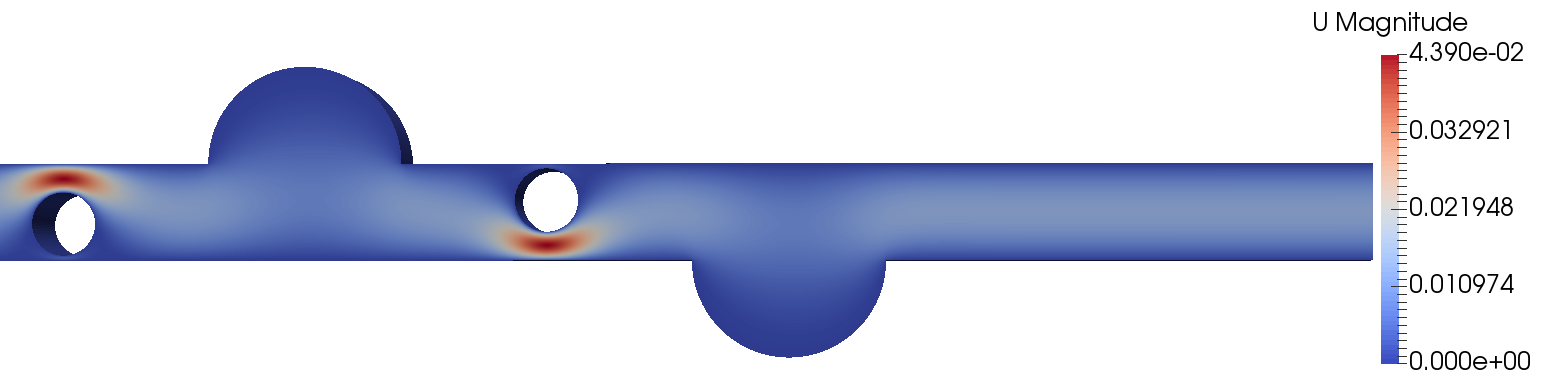}
}
\caption{Concentration (top) and velocity (bottom) profiles of the optimal cases.}
\label{fig1_resultado}
\end{center}
\end{figure}

\begin{figure}[h]
\begin{center}
\resizebox*{\textwidth}{!}{
\includegraphics[width=\textwidth]{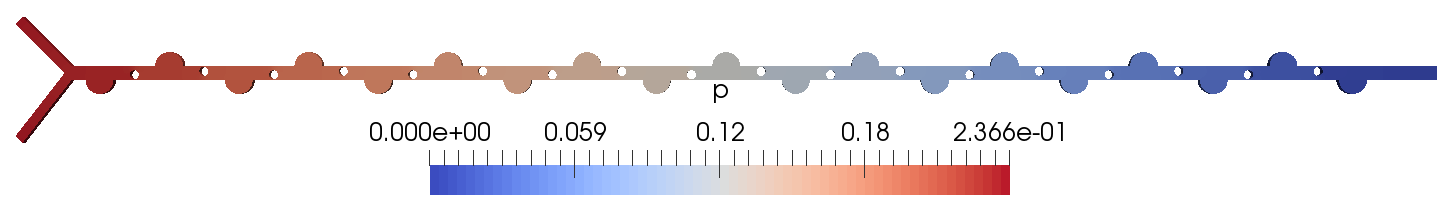}
}
\caption{Pressure profile of the optimal case.}
\label{fig3_resultado}
\end{center}
\end{figure}

Figures~\ref{fig1_resultado} and \ref{fig3_resultado} show the concentration, velocity and pressure profiles obtained through the simulation for the optimal case.

The errors obtained for $\varphi$ and $\Delta P$, respectively, for the optimized case were
$0.88\%$ and $3.60\%$.
Both the predicted values of $\varphi$ and $\Delta P$ are very accurate, specially for the optimum case.
The error for the maximum and minimum cases can be related to the limits of the multivariable method applied.

\section{Conclusions}

A variety of simulations were performed varying the diameter (OD) and the offset (OF) of circular obstructions along the channel of a micromixer.
The results obtained were used to train and test a neural network, obtaining errors of less than $1\%$ for values of $\varphi$ and less than $3\%$ for pressure drop. 

Global optimization was performed for the OD and OF configuration test set using genetic algorithm.
The results obtained point to the construction of an optimal micromixer with $OD = 131 \mu m$ and $OF = 10 \mu m$ for $Re \approx 1$.
This corresponds to a medium size obstacle that is close to the wall, which is compatible with previous results in the literature.

When verifying through the simulation the values of $\varphi$ and $\Delta P $ resulting from the optimization for the optimal case, we get errors of $0.88\%$ and $3.60\%$, respectively.
Besides the small erros, the values $\varphi = 57.975 \% $ and $\Delta P = 227.668 $ mPa of the simulated optimal geometry are even larger/smaller than the forecast of $57.47 \%$ and $ 297.46 $ mPa.

Finally, the use of Machine Learning, more specifically neural networks and genetic algorithm proved to be effective in the study of problems involving optimization.
It is important to emphasize that global optimization could be obtained by simulating 2000 cases using CFD. 
However, as each simulation takes around $4h$, the total time to guarantee the global optimization would be about 330 days. With this methodology, the whole process of producing the dataset, training and optimization takes 40 days. This procedure is a tremendous advantage for microfluidic optimization. %
It should also be emphasized that despite the case studied applied with a geometric optimization, the methodology will be the same for cases involving the optimization of any other parameters, such as input flow or diffusion coefficient, for example.

\bibliographystyle{elsarticle-num}

\end{document}